\RequirePackage{ifpdf}
\documentclass[hyper,letterpaper]{JHEP3}
\pdfoutput=1
\usepackage{amsmath,amssymb,amsfonts,bm,amscd,mathtools}
\usepackage{cite}
\usepackage{graphicx, caption, subcaption}
\usepackage{multirow}
\usepackage{verbatim}
\usepackage{appendix}
\usepackage{fancybox}
\usepackage{array}
\usepackage{url}
\usepackage{float}

\def\be{\begin{equation}}
\def\ee{\end{equation}}
\def\bea{\begin{eqnarray}}
\def\eea{\end{eqnarray}}

\DeclarePairedDelimiter\floor{\lfloor}{\rfloor}
\def\la{\alpha}

\title{In search of conformal theories}

\author{Abhijit Gadde
\\
School of Natural Sciences, Institute for Advanced Study, Princeton, NJ 08540, USA
}

\abstract{
The conformal crossing equation puts very stringent constraints on the conformal data. We formulate it in way that makes the conformal symmetry more transparent. This allows for generalization of the crossing equation to arbitrary Lie group $G$. Using the crossing equation for $SU(2)$ as a toy model, we find infinitely many solutions to the $G$-crossing equation. In particular, when $G$ is specialized to the conformal group $SO(d+1,1)$, we get infinitely many solutions to the conformal crossing equation. 
}

\begin{document}
\section{Introduction}
A $d$-dimensional euclidean conformal field theory is symmetric under the group $SO(d+1,1)$.
Formally it is defined in terms of the spectrum i.e. the set $S$ of conformal representations $r=[\Delta,\ell]$ where  $\Delta$ is the conformal dimension and $\ell = (\ell_1,\ldots, \ell_{\floor*{\frac{d}{2}}})$ is the spin and the structure constants $C_{r_1,r_2,r_3}$. 
The structure constant is a real valued function on ${\rm sym}^{\otimes 3}S$. The spectrum is constrained by the ``unitarity bound":
\be\label{unitarity-bound}
\Delta \in \Re, \quad \Delta \geq d/2-1\,\, {\rm for}\,\,\ell =(0,\ldots, 0),\quad \Delta \geq d/2+\ell'\,\, {\rm for}\,\,\ell = (0,\ldots,0,\ell') , \ldots
\ee
We have put the word unitarity in quotation marks because it does not refer to the unitary representations of the euclidean conformal group but rather to the unitary representations of the Lorentzian conformal group $SO(d,2)$ analytically continued to those of $SO(d+1,1)$. They are usually  the preferred representations because one is usually interested in the Lorentzian conformal theories from a physical point of view. We will refer to them as physical representations. This distinction between physical representations and euclidean unitary representations, to be described shortly, is important in the context of this paper.
The most nontrivial constraint on the conformal data comes from the  the crossing equation:
\be \label{crossing}
\sum_r C_{r_1,r_2,r}C_{r_3,r_4,r} F^{r_1,r_2,r_3,r_4}_{r}(u,v)=\sum_{r'} C_{r_1,r_3,r'}C_{r_2,r_4,r'} F^{r_1,r_3,r_2,r_4}_{r'}(v,u).
\ee
where $F$ are the so called conformal partial waves. The equation is so constraining,  it is generally believed to be sufficient.
The abstract approach to conformal field theory involves solving the crossing equation. 
This approach, appropriately called the conformal bootstrap, was pioneered in  \cite{Ferrara:1973yt, Polyakov:1974gs} and recently revived in \cite{Rattazzi:2008pe}.
Since then there has been a lot of success towards partially solving this equation by analytical as well as numerical methods, see \cite{Simmons-Duffin:2016wlq} for an example of the state of the art  results and  an extensive list of references.  

Our approach is to formulate the crossing  equation such that it depends only on the symmetry group $SO(d+1,1)$. From this point of view the crossing equation as presented above is somewhat unsatisfactory as it makes reference to the conformal cross-ratios  $(u,v)$ which are, in a sense, extraneous. In order to achieve the desired form, one would like to use orthogonality of conformal partial waves. So when integrated over $(u,v)$ with respect to appropriate measure one side of the equation gives Kronecker delta.
\be \label{crossing-kernel}
\sum_r C_{r_1,r_2,r}C_{r_3,r_4,r} \,W^{r_1,r_2,r_3,r_4;r,r'}=C_{r_1,r_3,r'}C_{r_2,r_4,r'} 
\ee
In this equation only representations of the conformal group appear. If one identifies the group theoretic properties of $W$ then the equation can be generalized to any group. 
Unfortunately, the  conformal partial waves for physical representations  do not obey an orthogonality relation. It is well known that the conformal partial waves are solutions to a differential equations \cite{Dolan:2003hv}. Such solutions enjoy orthogonality if the differential operator in question is hermitian. This is not the case for physical conformal representations. But for conformal representations $R=[\Delta,\ell]$ with $\Delta=d/2+ic, c\in {\mathbb R}$ the differential operator does become hermitian and associated conformal partial waves obey orthogonality. These representations are called the principal series representations and play an important role in the harmonic analysis on the group. In fact, the orthogonality of conformal partial waves of the principal series is closely related to the harmonic analysis.

How to hear the shape of a group? This is the question that is at the heart of harmonic analysis on $G$. To be precise, the space of normalizable functions on a topological group $G$ forms a representation under the left action of $G$ called the regular representation. How does it decompose into irreducible representations?  For compact groups, the complete basis of such  representations is a discrete set consisting of all irreducible representations while for non-compact groups, as it turns out, the basis consists of representations belonging to a continuous set and in some cases consists of an additional discrete set. 
The usefulness of this concept is immediately apparent in case of the simplest non-compact group ${\mathbb R}$. The harmonic analysis on ${\mathbb R}$ is simply the Fourier transform, clearly a useful notion. In the context of conformal theories, it is useful to decompose the 4-point functions into a \emph{complete basis} of conformal partial waves. The underlying problem is actually that of the harmonic analysis on the conformal group. 

The irreducible representations that appear in the decomposition of the regular representation of $G$ are known as tempered. By an old theorem of Harish-Chandra \cite{harish-chandra1970}, they are described as induced unitary representations of certain distinguished subgroups. We will identify these representations in case of the euclidean conformal group $SO(d+1,1)$. 
For even $d$ they fall in a continuous set known as the principal series while for odd $d$ they consist of the so called discrete series in addition to the principal one. 
\begin{itemize}
\item Principal series representations: $[\Delta,\ell]$ with $\Delta=\frac{d}{2}+ic,\,c\in \Re$.
\item Discrete series representations (only for odd $d$):  $\Delta=\frac{d}{2}+{\mathbb Z}_+$ 
\end{itemize}
The physical representations \emph{i.e.} the representations satisfying the unitarity bound \eqref{unitarity-bound} \emph{do not} appear in the decomposition of the regular representation. There are other unitary representations beside the tempered ones \emph{e.g.} the complementary series: $\Delta\in (0,d)$ but they will not play any role in our discussion. In the rest of the paper we will only consider the case of conformal theory in even dimensions, so the discrete series will also not appear in the rest of the paper. However, the discussion can be generalized to odd dimensions by adding the discrete series to various formulas. We denote the tempered representations by capital letter $R$ to distinguish them from the physical representations $r$.

We study the Clebsch-Gordan decomposition of the tensor product of two tempered representations into irreducible ones. Only the tempered representations appear in the decomposition. The dynamics must be consistent with kinematics. In particular, the 3-point function is non-zero only if it is allowed by the fusion rules. 
The associativity of 3-point functions results in a crossing equation very similar to \eqref{crossing} except that the conformal representations appearing are not physical but rather are tempered. Correspondingly the sum is replaced by an integral over the principal series with appropriate measure (along with sum over spins and, in the case of odd $d$, sum over the discrete series). This new ``euclidean" crossing equation can be rightly thought to define a \emph{unitary euclidean conformal theory}. Clearly, these theories are different from the physical conformal theories that  are unitary in Lorentzian signature. Nevertheless, it seems plausible to generate the solution to the physical crossing equation \eqref{crossing} given an analytic solution to the euclidean crossing equation by way of residue integration. This is achieved by deforming the contour of integration along the principal series representations: $\Delta=\frac{d}{2}+ic$, and picking up poles on the positive real axis corresponding to physical representations. We will elaborate on this later in the paper.
  
The advantage of working with tempered representations is that one can now use the orthogonality of the conformal partial waves to write the crossing equation as \eqref{crossing-kernel} with, by now familiar, substitution of the sum by principal series integral. The object $W$ appearing in the equation is nothing but the  Racah coefficient\footnote{Also known as the $6j$-symbol when normalized more symmetrically.} of the euclidean conformal group $SO(d+1,1)$. This form of the equation admits generalization to any group $G$. We will find infinitely many analytical solutions to this generalized $G$-crossing equation. In this effort the crossing equation for the simplest nontrivial Lie group $SU(2)$, will serve as our  main guide.

The rest of the paper is organized in the following way. In section 2, we will review the relevant aspects of the representation theory of $SO(d+1,1)$. This includes the construction of tempered representations via induction and a discussion of unitarity. 
We will also discuss the Clebsch-Gordan decomposition of the direct product of two representations. The crossing equation is formulated as a consequence of the associativity of direct product. This formulation extends to a general group $G$. 
In section 3 we will find the solutions of the $G$-crossing equation. One of the solutions to the $SU(2)$-crossing equation is already known in the literature. It is obtained as a result of the associativity of the so called ``symplecton polynomials". In the modern language it is  understood as the associativity of the operator ring of a certain topological quantum mechanical system with $SU(2)$ action on the phase space. Inspired by the underlying algebra we will construct infinitely many solutions for the general crossing equation in terms of the Racah coefficient itself. The proposed solution satisfies the crossing equation thanks to the Biedenharn-Elliot identity  also known as  the pentagon identity of  the Racah coefficient. The solutions are in one to one correspondence with  representations of $G$. We end with a discussion of some of the new directions opened as a result of this work. As Racah coefficient plays a prominent role in formulation of the crossing equation as well as in its solutions, we have added an appendix containing its definition and properties.

\section{Conformal representation theory and the crossing equation}\label{conformal-rep}
This section is essentially a quick review of conformal representation theory based on the excellent reference \cite{Dobrev:1977qv}.
A discussion of conformal theory usually begins with locality, in particular with the definition of local operators defined in an ambient $d$-dimensional space. 
For our purposes, however,  it is more useful to assign primary importance to the symmetry group. In the case of $d$-dimensional euclidean theory, the symmetry group is $SO(d+1,1)$. The ambient space of the theory would be emergent.
We ask how does the left-regular representation of the conformal group decompose into its irreducible representations. 

An element of $SO(d+1,1)$ is written uniquely as $g=kan$ where $k$ is an element of the maximal compact subgroup $SO(d+1)$, $a$ is scale transformation and $n$ is a special conformal transformation. For general semisimple groups, this decomposition is known as the Iwasawa decomposition and is expressed as $G=KAN$. Thanks to Harish-Chandra \cite{harish-chandra1970}, it is known that the irreducible representations appearing in the harmonic decomposition are the unitary ones induced by the subgroup $MAN$ where $M$ is the commutant of $A$ inside $K$ and that the inducing representation transforms trivially under $N$. In the case of the conformal group $M=SO(d)$. Such a representation is a function on the coset $G/MAN$ labeled by its transformation under $A$ i.e.  the conformal dimension $\Delta$ and under $M$ i.e. the spin $\ell$.  The coset $G/MAN$ is nothing but the familiar conformal compactification of $d$-dimensional euclidean space. This is how the coordinate space emerges from abstract symmetry considerations.

The above discussion can be phrased equivalently and perhaps in a more familiar fashion by starting from the euclidean space and considering representations of the little group $MAN$ that are invariant under $N$. The little group is the group of transformations that fixes a point (where a local operator is supported). The advantage of the more formal approach is that it can be generalized to other non-compact groups. As we seek to put the conformal theory in a group theoretic framework, this is helpful for applications that we have in mind.
To summarize, the representations of the conformal group appearing in the harmonic analysis are vectors on ${\mathbb R}^d$ labeled by their representations $[\Delta, \ell]$ under the little group. 

\subsection{Unitarity and completeness}

From the point of view of harmonic analysis, we are interested in the induced representations of $MAN$ that are unitary. Before discussing unitarity let us observe that an invariant inner product can be defined between a representation $[\Delta,\ell]$ and $[d-\Delta,\ell]$. Letting $f_1$ and $f_2$ be certain vectors from these representations respectively,
\be
\langle f_2|f_1 \rangle = \int dx\,\, {f}_2(x) \cdot f_1(x).
\ee
Here $\cdot$ stands for inner product in the finite dimensional representation of $SO(d)$. It is not difficult to verify that the inner product defined above is indeed invariant \cite{Dobrev:1977qv}. For example, under scale transformation $x\to \lambda^{-1} x$: $dx^d \to \lambda^{-d} dx^d, f_1\to \lambda^\Delta f_1,  f_2\to \lambda^{d-\Delta} f_2$ which keeps the integral invariant. A similar argument works for showing invariance under special conformal transformation. The inner product defines a map from the representation $R\equiv [\Delta,\ell]$ to ${\mathbb R}$. This allows us to interpret the representation $\tilde R\equiv [d-\Delta,\ell]$ as the dual of representation~$R$. In conformal field theory literature the representation $\tilde R$ is also known as the shadow of representation $R$.
Note that for a given representation, the shadow can be constructed through convolution. As an example, if $f(x)$ is a scalar representation with conformal dimension $\Delta$ then
\be\label{convolution}
{\tilde f}(x') = {\cal N}_{\Delta}\int  d^d x |x-x'|^{2(\Delta-d)} f(x),\qquad {\cal N}_{\Delta}=\frac{\pi^d \Gamma(\Delta-\frac{d}{2})\Gamma(\frac{d}{2}-\Delta)}{\Gamma(\Delta)\Gamma(d-\Delta)}.
\ee
The normalization ${\cal N}_{\Delta}$ is fixed by requiring the 2-point function of $f(x)$ as well as $\tilde f(x)$ to be unity:
\be\label{normalization}
\langle f(x) f(x')\rangle =|x-x'|^{-2\Delta}, \qquad \langle {\tilde f}(x) {\tilde f}(x')\rangle =|x-x'|^{2(\Delta-d)}. 
\ee

To identify unitary representations we simply need notice that for $\Delta = \frac{d}{2}+ic, c\in {\mathbb R}$, the dual of the representation $[\Delta,\ell]$ is its complex conjugate.  Such representations  form the so called \emph{principal series} and play an important role in the harmonic analysis. In fact, for even $d$, the principal series representations are the only representations that appear in the harmonic analysis. For odd $d$, in addition to the principal series representations, the so called discrete series representations also appear in the harmonic analysis. This is due to the existence of a compact Cartan subgroup of the conformal group \cite{harish-chandra1970}. In this paper we will focus on the case of even $d$. Most of our results can be generalized to odd $d$ with minor modifications.
On the other hand, the familiar physical representations with real $\Delta$ satisfying the unitarity bound are not very natural from the point of the unitarity as defined above. As remarked earlier, they are analytical continuation of the unitary representations of the 
Lorentzian conformal group $SO(d,2)$.

What is the completeness relation obeyed by the representations of the conformal group? To see what we mean exactly, consider a simple case of the translation group ${\mathbb R}$. A representation of the translational group is given by the momentum eigenstate $e^{2\pi i px}$ where $p$ is a complex number. Given an normalizable function $f(x)$, it can be decomposed into translation eigenstates with coefficients ${\hat f}(p)$. We know that the Fourier transform ${\hat f}(p)$ is supported only on the real axis. In other words, the representations with \emph{real} $p$ form a complete basis for the space of  normalizable functions on ${\mathbb R}$. This completeness is sometimes expressed as,
\be
f(0)=\int_{\Im p=0} dp {\hat f}(p),  \qquad {\rm where}\qquad {\hat f}(p)\equiv \int_{-\infty}^{\infty} dx\, f(x)e^{2\pi ipx}.
\ee
Harmonic analysis is essentially a Fourier transform on the group manifold. As discussed above, a normalizable function on $SO(d+1,1)$ for even $d$ can be decomposed into the principal series representations. Let $R=[\frac{d}{2}+ic, \ell]$ be a principal series representation. It plays the role of the Fourier basis. The above completeness relation is generalized to,
\be\label{fourier-group}
f(1)=\int dR\, {\rm Tr}[{\hat f}(R)],\quad {\rm where}\,\, {\hat f}(R)\equiv \int dg \,f(g)R(g),\,\,  \int dR\equiv  \sum_{\ell}\int_{\frac{d}{2}-i\infty}^{\frac{d}{2}+i\infty} \rho_\ell(\Delta)d\Delta
\ee
Here $dg$ is the Haar measure on the group. The sum $\sum_\ell$ is over all the spins. The function $\rho_\ell(\Delta)$ is known as the Plancherel weight and is known to be equal to the normalization constant in the integration kernel relating representation $[\Delta,\ell]$ to its shadow  \cite{Knapp-Stein, Dobrev:1977qv}. For scalars, it is equal to ${\cal N}_{\Delta}$ in equation \eqref{convolution}.
For odd dimensions the integral over the principal series is supplemented with a sum over the discrete series. In odd dimensions, the Plancherel weight has poles, the discrete series of representations lives at those points and essentially serves to cancel the ``fake" poles coming from the measure.

\subsection{Clebsh-Gordan decomposition}
Conformal theories are usually formulated in terms of the operator product expansion. The product of local operators at two distinct points is presented in terms of a single local operator insertion. 
\be\label{OPE}
O_1(x_1)O_2(x_2) = \sum_O K_{O_1,O_2,O} \, O(x) +{\rm desc.}
\ee
The operators are normalized using their two point function. The tower of descendants denoted as desc. is completely fixed by symmetries. 
The constant $K$ is a dynamical coefficient known as the three point function coefficient or the structure constant. Apart from the spectrum, $K$ is the only dynamical data defining the conformal theory. It is constrained by the crossing equation \eqref{crossing} which follows from the associativity of the operator product expansion. Each local operator is a representation of the conformal group. It is then natural to look at this expansion from the point of view of representation theory i.e. to ask how is the operator product expansion consistent with the representation ring of the conformal group. This question can be answered for the principal series representations. 

The direct product of two principal series representations can be decomposed into irreducible representations. As it turns out only the principal series representations appear in this decomposition \cite{Dobrev:1977qv}. In other words, the principal series representation are closed under Clebsch-Gordan decomposition. A principal series representation is a function on ${\mathbb R}^d$, $f_{R=[\Delta, \ell]}$ labeled by the conformal dimension $\Delta$ and spin $\ell$. The vector in the direct product of two such representations is a bilocal function $ f(x_1,x_2)\equiv f_{R_1}(x_1)f_{R_2}(x_2)$. On general grounds, the Clebsch-Gordan takes the form,
\be\label{CG-conformal}
f(x_1,x_2)=\int dR\int dx \,\, C^{R_1,R_2,\tilde R}(x_1,x_2,x) f_{R}(x).
\ee
The integration measure $dR$ is defined in \eqref{fourier-group}. The kernel $C$ is the Clebsch-Gordan coefficient. Let us emphasis that so far this is a purely group theoretic statement and does not include the dynamical structure constant $K$. Comparing with the Clebsch-Gordan decomposition for a compact group, say $SU(2)$:
\be
|j_1,m_2\rangle \otimes |j_2,m_2\rangle = \sum_{j,m} C^{j_1,j_2,j}_{m_1,m_2,m}|j,m\rangle,
\ee
we see that the coordinates $x_i$ are analogous to the magnetic quantum number $m_i$. The explicit form of the integration kernel is determined by matching conformal transformation properties on both side of equation \eqref{CG-conformal}.
Surprisingly, this exercise turns out to be easier than determining the Clebsch-Gordan coefficients for the compact group. As an example, for scalar operators, 
\be\label{3-pt-function}
C^{R_1,R_2,R_3}(x_1,x_2,x_3) =\frac{\cal C}{|x_{12}|^{\Delta_1+\Delta_2-\Delta_3}|x_{23}|^{\Delta_2+\Delta_3-\Delta_1}|x_{31}|^{\Delta_3+\Delta_1-\Delta_2}}, \quad x_{ij}\equiv x_i-x_j.
\ee
Normalization constant $\cal C$ is fixed from the normalization of the 2-point function  \eqref{normalization}. 
It is clear from equation \eqref{3-pt-function} that the Clebsch-Gordan coefficient for the conformal group has the same form as the familiar 3-point function. Indeed the operator product expansion \eqref{OPE} can be expressed as an integral over space with a Kernel that is the three point function instead of as sum over descendents. This is described in \cite{Czech:2016xec}. The integral form of the operator product expansion is termed the OPE block.

The Clebsch-Gordan coefficient for decomposing the direct product of representations $R_1$ and $R_2$ into $R_3$ is nonzero if the 3-point function of $R_1,R_2$ and ${\tilde R}_3$ is nonzero. A configuration of three points on the sphere breaks the conformal symmetry to $SO(d-1)$. So the three point function is nonzero only if the tensor product of spins $\ell_1,\ell_2$ and $\ell_3$ admit an $SO(d-1)$ singlet.
The Clebsch-Gordan coefficients also enjoy an orthogonality relation. It is most familiar in the case of $SU(2)$,
\be\label{su2-ortho}
\sum_{m_1,m_2} C^{j,j_1,j_2}_{m,m_1,m_2}C^{j_1,j_2,j'}_{m_1,m_2,m'} = \delta_{j,j'}\delta_{m,m'}.
\ee
Similarly for the conformal group \cite{Dobrev:1977qv},
\bea\label{conformal-ortho}
\int dx_1dx_2 C^{R,R_1,R_2}(x,x_1,x_2) C^{{\tilde R}_1,{\tilde R}_2,R'}(x_1,x_2,x')&=&\delta(x-x') \delta_{R,{\tilde R}'}+G(x-x') \delta_{R,R'}\nonumber\\
\delta_{R,R'}&\equiv & \delta_{\ell \ell'} \delta(\Delta-\Delta')\frac{2\pi}{\rho_\ell(\Delta)}.
\eea
Here $\rho_\ell(\Delta)$ is the Plancherel weight defined in equation \eqref{fourier-group} and  $G(x-x')$ is the two point function of two local operators with representation $R$. For example, if $R=[\Delta,0]$ then $G=|x-x'|^{-\Delta}$.
This equation will be useful later in the paper.

\begin{figure}[t]
\centering
\includegraphics[scale=0.3]{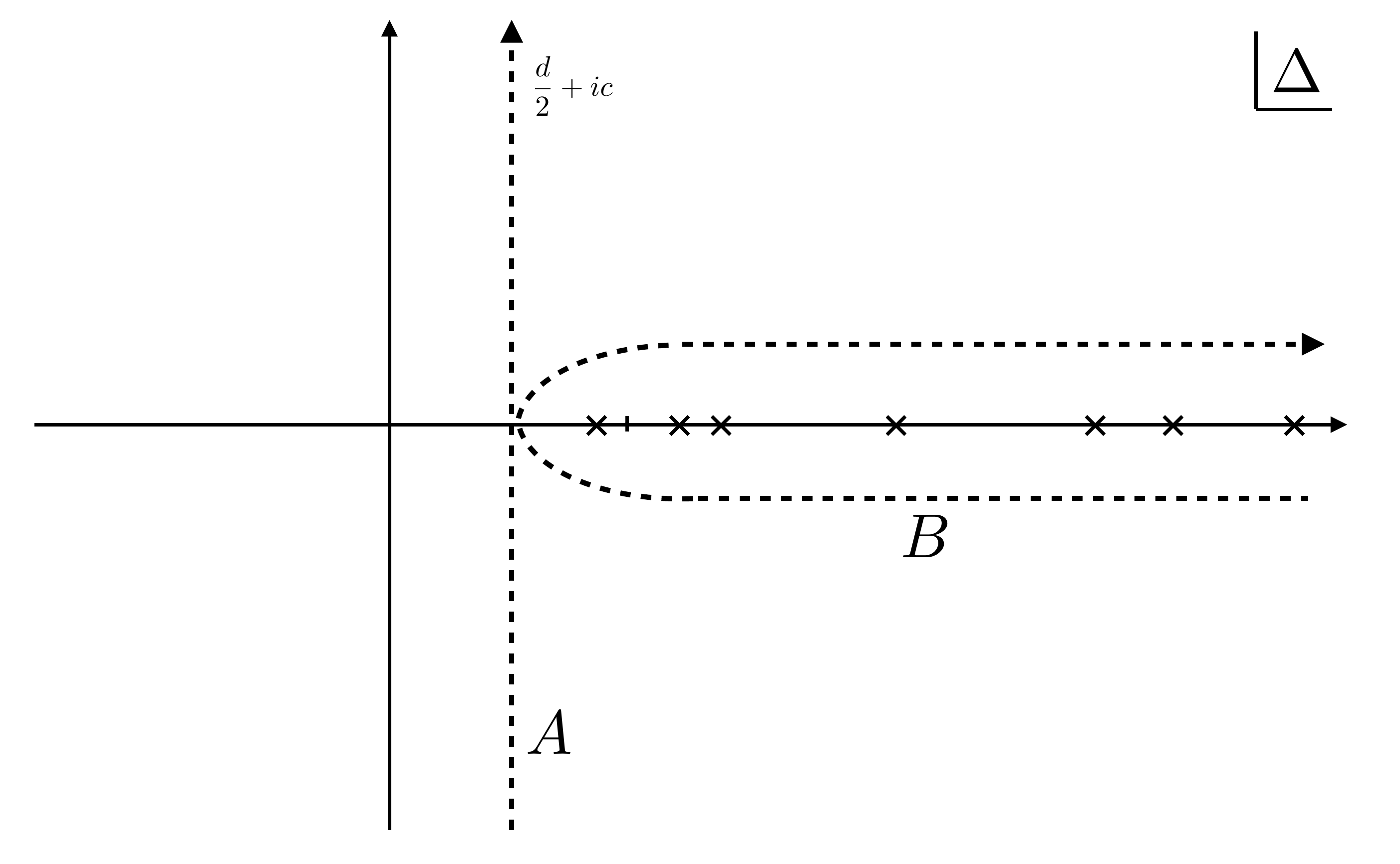}
\caption{The integral over the principal series representations is represented by the contour A. It can be expressed as a sum over physical representations corresponding to the poles on the real axis given that the contour can be deformed to B without picking any additional complex poles.}
\label{integral-to-sum}
\end{figure}

In a unitary euclidean conformal theory, the operator product expansion is best expressed in a way that makes the  representation ring \eqref{CG-conformal} manifest. The only additional piece required is the structure constant $K$. It is incorporated as follows,
\be\label{CG-OPE}
\boxed{
O_{R_1}(x_1)O_{R_2}(x_2)=\int dR\int dx \,\, K_{R_1,R_2,R} \,C^{R_1,R_2,R}(x_1,x_2,x) O_{\tilde R}(x).
}
\ee
Interestingly this suggests an extension of the notion of operator product expansion to  compact groups as well as to any other non-compact groups. Taking the example of $SU(2)$, a suitable 
``operator product expansion" would be,
\be\label{su2-ope}
O_{j_1,m_1} O_{j_2,m_2} = \sum_{j,m} K_{j_1,j_2,j}C^{j_1,j_2,j}_{m_1,m_2,m}O_{j,m}.
\ee
The structure constants $K$ are symmetric and the operators $O_{j,m}$ are normalized such that $K_{j,j,0}=1$. Given that the group theory allows a nonzero 1-point only for the trivial representation, normalizing it to unity $\langle O_{0,0}\rangle=1$, we get $\langle O_{j_1,m_1} O_{j_2,m_2} \rangle =\delta_{j_1,j_2}\delta_{m_1+m_2,0}C^{j,j,0}_{m,-m,0}$.
The constraint on $K$ comes from the associativity of the operator product expansion. For euclidean conformal group, the associativity equation is
\bea\label{assoc}
&& \int dR K_{R_1,R_2,R}K_{{\tilde R},R_3,R_4} \int dx \, C^{R_1,R_2,R}(x_1,x_2,x)C^{{\tilde R},R_3,R_4}(x, x_3,x_4)=\nonumber\\
&&\int dR' K_{R_2,R_3,R'}K_{R_1,{\tilde R}',R_4}  \int dx \, C^{R_2,R_3,R'}(x_2,x_3,x')C^{{R_1,{\tilde R}',R_4}}(x_1, x',x_4).
\eea
It is known that the position space integral on each side of the integral is proportional to the sum of conformal blocks $g_R$ and $g_{\tilde R}$ for representation $R$ and its shadow ${\tilde R}$ \cite{SimmonsDuffin:2012uy}. For scalar external operators,
\bea
&&\int dx \, C^{R_1,R_2,R}(x_1,x_2,x)C^{{\tilde R},R_3,R_4}(x, x_3,x_4)=\\
&& |x_{12}|^{\Delta_1+\Delta_2-\Delta}|x_{34}|^{\Delta_3+\Delta_4-\Delta}(g_{R}(u,v)+g_{\tilde R}(u,v)),
\qquad u=\frac{|x_{13}x_{24}|}{|x_{12}x_{34}|},\, v=\frac{|x_{14}x_{23}|}{|x_{12}x_{34}|}\nonumber.
\eea
It is clear that the equation \eqref{assoc} is very similar to the crossing equation \eqref{crossing}, the main difference is that the physical representations have been replaced by principal series representations and correspondingly the sum is replaced by an integral.  
In fact they are related even more  intimately. If the structure constants have appropriately slow asymptotic growth then due to the exponentially decaying behavior of the conformal block $g_R$ towards positive real infinity, the contour for the first term can be deformed towards the positive real axis as shown schematically in figure \ref{integral-to-sum}.
The shadow block $g_{\tilde R}$ decays exponentially towards negative real infinity so its contour  can be deformed towards negative real axis\footnote{We thank Balt van Rees for discussion on this issue.}. Thus the integral  over principal series is converted to a sum over physical representations. Here we have assumed that the poles of the integral lie only on the real line. The upshot  is that if one finds an analytic solution to equation \eqref{assoc} with appropriate properties then one can build the solution of the physical crossing equation \eqref{crossing}. One subtle obstruction to this argument is that, say on the left hand side of the equation,  the product of structure constants that appear are $K_{R_1,R_2,R}K_{{\tilde R},R_3,R_4}$ and not $K_{R_1,R_2,R}K_{{R},R_3,R_4}$; similarly on the right. It is the later combination that would produce the desired equation. So in addition to their symmetry under all permutations of their labels, we also require the structure constants to be shadow symmetric i.e.
\be\label{shadow-symmetry}
K_{R_1,R_2,R_3}= K_{{\tilde R}_1,R_2,{R}_3}.
\ee
With this condition, the two terms in the integral consisting of $g_{R}$ and $g_{\tilde R}$ give the same contribution.

\begin{figure}[t]
\centering
\includegraphics[scale=0.3]{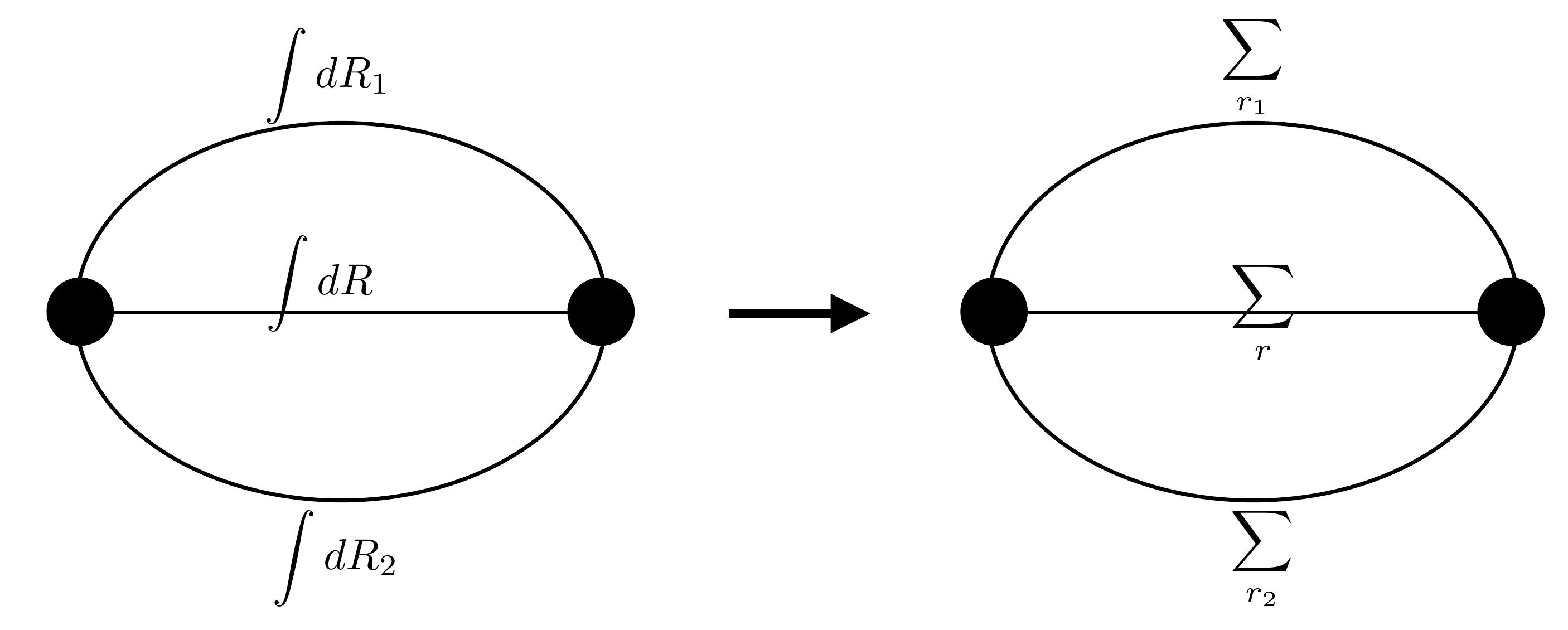}
\caption{Converting the principal series representations to physical representations.}
\label{simultaneous}
\end{figure}

The above argument is for when the external representations $R_1,\ldots, R_4$ are kept fixed. As the external representations are also principal series representations, it is not sufficient to just convert the intermediate representations to physical ones. We propose the following method to change all the representations from principal series to physical set together. Take $R_1=R_3$ and $R_2=R_4$. The poles of the integral, say on the left hand side of equation \eqref{assoc}, are then determined in terms of principal series representations $R_1$ and $R_2$. This gives the equation,
\be
R=f_i(R_1,R_2).
\ee
Here $i$ labels the poles. As the structure constant is symmetric in all three representation labels, we symmetrize the above equation to obtain two additional equations. The physical representations then live at the intersection of three divisors $f_i(R_1,R_2), f_j(R,R_1)$ and $f_k(R_2,R)$. This procedure is summarized in figure \ref{simultaneous}. We have denoted the Clebsch-Gordan coefficient $C^{R_1,R_2,R}$ by a trivalent vertex. The accompanying factor of $K$ is denoted as a black dot.

Before embarking on finding the solutions to equation \eqref{assoc}, it is instructive to consider the more familiar case of $SU(2)$. The associativity constraint coming from the product \eqref{su2-ope}.
\be\label{su2-crossing}
\sum_{j,m} K_{j_1,j_2,j}K_{j,j_3,j_4} C^{j_1,j_2,j}_{m_1,m_2,m}C^{j,j_3,j_4}_{m,m_3,m_4} = \sum_{j',m'} K_{j_2,j_3,j'}K_{j_1,j',j_4} C^{j_2,j_3,j'}_{m_2,m_3,m'}C^{j_1,j',j_4}_{m_1,m',m_4}.
\ee
This equation is simplified further using what is known as  the Racah coefficient. We have defined the Racah coefficient in appendix \ref{6j-symbol}. It is the recoupling coefficient for three angular momenta.
Even though it is defined in an algebraic fashion, for our current purposes it can be thought of as the coefficient that expresses the product of two Clebsch-Gordan coefficients in one channel in terms of linear combination of the product of Clebsch-Gordan coefficients in the other channel, see equation  \eqref{CG-Racah}.
\be\label{su2-6j}
C^{j_1,j_2,j}_{m_1,m_2,m}C^{j,j_3,j_4}_{m,m_3,m_4} = \sum_{j'} \sqrt{(2j+1)(2j'+1)}
W(j_1,j_2,j_3,j_4;j,j')
\, C^{j_2,j_3,j'}_{m_2,m_3,m'}C^{j_1,j',j_4}_{m_1,m',m_4}.
\ee
The  quantity $W$ is the Racah coefficient\footnote{A more symmetric object is known as the $6j$ symbol, it is defined as $(-1)^{j_1+j_2j_3+j_4}W$. It enjoys a group of symmetries  consisting of  the symmetries of a tetrahedron.}. Substituting this expression in equation \eqref{su2-crossing} and comparing each term in the $j$ sum we get an equation that doesn't depend on the magnetic quantum number $m_i$. In comparing individual terms in the $j$ sum we make use of the orthogonality condition \eqref{su2-ortho}.
\be\label{su2-6j-crossing}
K_{j_2,j_3,j'}K_{j_1,j',j_4}  = \sum_{j'} \sqrt{(2j+1)(2j'+1)}
W(j_1,j_2,j_3,j_4;j,j')\,  K_{j_1,j_2,j}K_{j,j_3,j_4}.
\ee
This form of the associativity constraint is more invariant than \eqref{su2-crossing} as it does not refer to the individual vectors in the representations. It is easy to generalize to other Lie groups especially to non-compact Lie groups. Due to its abstract origin, the Racah coefficient can defined for any group. The representations that would appear are the ones arising in the harmonic analysis.
In the case of compact groups this set  consists of all irreducible representations while for non-compact groups it consists of certain types of induced representations described in section \ref{conformal-rep}. We refer to the generalization of  the associativity equation \eqref{su2-6j-crossing} or equivalently  \eqref{su2-crossing} to Lie group $G$ as $G$-associativity equation or $G$-crossing equation. For even $d$, the $SO(d+1,1)$-associativity equation is,
\be\label{conformal-6j-crossing}
\boxed{
K_{R_1,R_2,R}K_{R_3,R_4,R} = \int dR' \,
W(R_1,R_2,R_3,R_4;R,R')
\, K_{R_1,R_3,R'}K_{R_2,R_4,R'}.
}
\ee
In this equation the function $W$ is the Racah coefficient for the conformal group and the integral is over the principal series. This equation is equivalent to \eqref{assoc} and the  comments below the latter about contour deformation and changing the principal series integral to a sum over physical representations are equally applicable to this equation i.e. a solution to \eqref{conformal-6j-crossing} with appropriate analytic and asymptotic behavior leads to a physical solution to the conformal crossing equation.
From the relationship between equations \eqref{conformal-6j-crossing} and \eqref{assoc} we see  that the ambient space does not play a profound role in defining the conformal theory but can be thought of as merely an artifact of the group theory.
 In the remainder of the paper we find explicit solutions to $G$-associativity equation for any $G$.

\section{Solutions to generalized crossing equation}
Let us first consider the $SU(2)$-associativity equation \eqref{su2-6j-crossing}. 
The advantage of working with a compact group is that one has to deal with a discrete set of representations rather than a continuous one. Interestingly one solution to the $SU(2)$ associativity  is already known in the literature \cite{Biedenharn:1981er, Biedenharn:1981fq}:
\be\label{symplecton}
K_{j_1,j_2,j_3}= \frac{1}{[(2j_1+1)(2j_2+1)(2j_3+1)]^{\frac14}}\Big[\frac{(j_1+j_2+j_3+1)!}{(j_1+j_2-j_3)!(j_1-j_2+j_3)!(-j_1+j_2+j_3)!}\Big]^{\frac12}
\ee
It is obtained using the  associativity of multiplication of the so called ``symplecton polynomials" \cite{BIEDENHARN1971459, Mukunda1974}. From the modern perspective, the  idea  is to consider the quantization of  $\mathbb{CP}^1$ and study the associated ring of operators. The phase space enjoys an action of $SU(2)$ and so do the functions on the phase space.  After quantization  these functions  become operators acting on the Hilbert space. 
They  form a vector space. 
If the quantization is performed covariantly, the vector space of quantum mechanical operators also inherits the $SU(2)$ action.   We can organize the operators into irreducible representations of this action.
 Let $P^j_m$ to be an operator belonging the representation $j$ of $SU(2)$ with charge  $m$ under the Cartan generator. Multiplying together two vectors from different representations and decomposing the resulting operator into irreducible representations we expect
\be\label{operator-mult}
P^{j_1}_{m_1} \cdot P^{j_2}_{m_2}= \sum_{j,m} \,{K}_{j_1,j_2,j} C^{j_1,j_2,j}_{m_1,m_2,m} \, P^j_m.
\ee
Here $\cdot$ stands for operator multiplication. The $m$-dependence of the right hand side is uniquely fixed by group theory while  $K$ is an undetermined constant allowed by  the symmetry. Lo and behold, the equation \eqref{operator-mult} is exactly the same as equation \eqref{su2-ope}. 
We expect the operator multiplication to be associative. This associativity constraint is the same as  equation \eqref{su2-crossing}. In this way the $\mathbb{CP}^1$ quantum mechanical system  has generated a solution to the $SU(2)$-associativity equation. The authors of \cite{BIEDENHARN1971459, Mukunda1974} find the solution \eqref{symplecton} by explicitly constructing operators $P^j_m$. For similarly constructed solutions to the associativity equation for the quantum group $SU(2)_q$ see \cite{Nomura}.

From this discussion it is clear that quantization of the classical phase space admitting $G$-symplectomorphism can be use to generate a solution to the $G$-associativity equation. In general the $G$-action will not be carried over to the Hilbert space but only to the space of operators but if the coordinate space  itself enjoys a $G$-action then the group will also act on the Hilbert space. In this case we can find the solution to  $G$-associativity explicitly.
As we will describe momentarily, the idea is in fact more basic and can be formulated in an algebraic way without relying on quantum mechanics. Identifying the underlying algebraic structure will get us infinitely many solutions in closed form. They will be in one to one correspondence with representations of $G$. We will describe the construction for $SU(2)$, it admits straightforward generalization to any Lie group including the conformal group.

\begin{figure*}[t]
    \centering
    \begin{subfigure}[h]{0.5\textwidth}
        \centering
        \vspace{2.1cm}
        \includegraphics[scale=0.3]{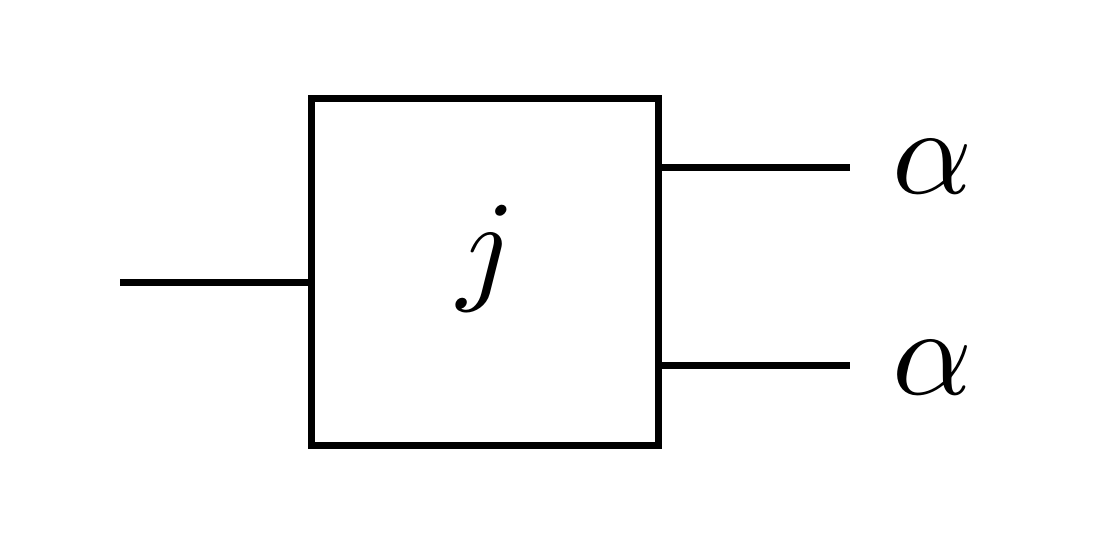}
        \vspace{2cm}
        \caption{The operator in representation $j$ is represented graphically. The two lines denote the vector spaces $\la$. When the operator is represented as a matrix, they also represent its two indices.}\label{operator-fig}
    \end{subfigure}%
    ~ 
    \begin{subfigure}[h]{0.5\textwidth}
        \centering
        \includegraphics[scale=0.3]{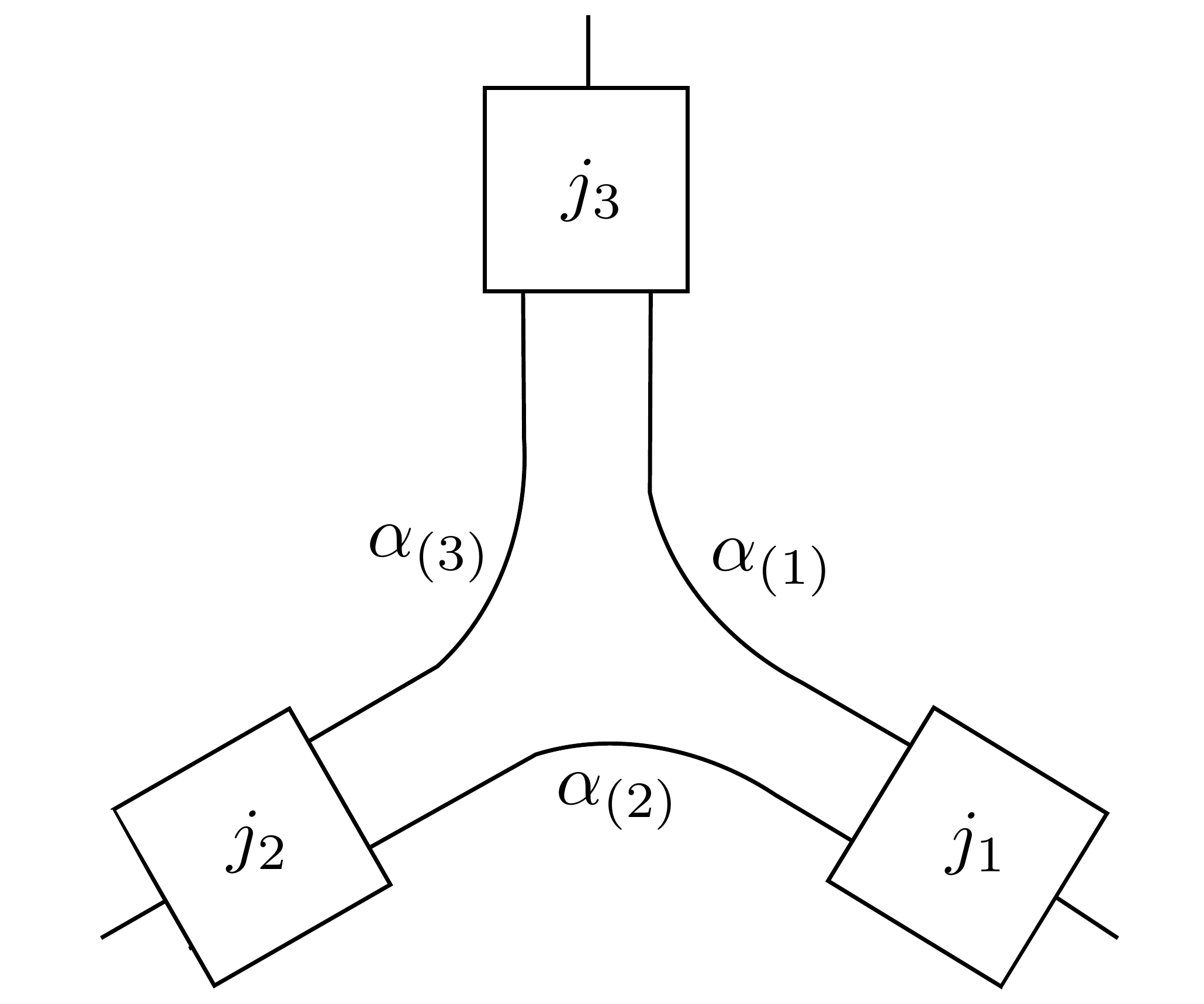}
        \caption{The operator product algebra. This figure denotes the $j_3$ component in the multiplication of representation $j_1$ and $j_2$ respectively. The subscript on $\la$ distinguishes the three isomorphic vector spaces.}\label{operator-mult-fig}
    \end{subfigure}
    \caption{Graphical representation of the operators and their product algebra.}
\end{figure*}

Let the Hilbert space form a representation $\la$ of $SU(2)$. The linear operators acting on this space belong to the representation $\la\otimes \la^*$ where $\la^*$ is the dual  to $\la$. In the case of $SU(2)$, $\la^*=\la$. We organize these operators according to their irreducible representations $j$ i.e. $\la\otimes \la^*=\oplus_{j=0}^{\la} j$. 
In figure \ref{operator-fig} we have denoted the operator $O_j$ in representation $j$ graphically. 
The two directed lines represent its matrix indices valued in  representations $\la$. Multiplication of two operators is obtained by contracting the matrix indices appropriately. Graphically, the index contraction is represented by  joining one of the ``internal" $\la$ lines of the two operators.  
The component of the operator $O_{j_3}$ in irreducible representation $j_3$ in the multiplication of operators $O_{j_1}$ and $O_{j_2}$ is denoted in figure \ref{operator-mult-fig}. More specifically, if we pick  vectors $m_i$ belonging to representation $j_i$ then the figure \ref{operator-mult-fig} refers to the coefficient $K_{j_1,j_2,j_3}C^{j_1,j_2,j_3}_{m_1,m_2,m_3}$ that appears on the right hand side of equation \eqref{operator-mult}.
If the ``internal" representation $\la$ is irreducible one can compute the structure constants $K_{j_1,j_2,j_3}$ explicitly. Let us use the notation $\la^{(i)}$ to distinguish the three irreducible isomorphic representations  appearing in figure \ref{operator-mult-fig} and let $n^{(i)}$ index their vectors. Then,
\be
j_1\in \la_{(1)}\otimes \la_{(2)},\qquad j_2\in \la_{(2)}\otimes \la_{(3)},\qquad j_3\in \la_{(3)}\otimes \la_{(1)}.
\ee
Correspondingly, their vectors are related by Clebsch-Gordan coefficients,
\be
|j_1,m_1\rangle = \sum_{n_{(1)},n_{(2)}} C^{j_1,\la_{(1)}, \la_{(2)}}_{m_1,n_{(1)},n_{(2)}} |\la_{(1)}, n_{(1)}\rangle |\la_{(2)}, n_{(2)}\rangle , \, \qquad\ldots .\nonumber
\ee
The quantity denoted in figure \ref{operator-mult-fig} is a combination of product of three Clebsch-Gordan coefficients.
\be
\sum_{n_{(1)},n_{(2)},n_{(3)}} C^{\la_{(2)}, j_1,\la_{(1)}}_{n_{(2)}, m_1,n_{(1)}} 
C^{\la_{(3)}, j_3,\la_{(1)}}_{n_{(3)}, m_3,n_{(1)}}
C^{\la_{(3)},j_2,\la_{(2)}}_{n_{(3)}, m_2,n_{(2)}} = K_{j_1,j_2,j_3}C^{j_1,j_2,j_3}_{m_1,m_2,m_3}.
\ee
Thanks to the definition of the Racah coefficient \eqref{su2-6j} and orthogonality \eqref{su2-ortho}, the left hand side can be evaluated in terms of the Racah coefficient, see equation \eqref{CG-Racah-relation}. It is indeed proportional to the Clebsch-Gordan coefficient as on the right hand side. The proportionality constant, appropriately normalized is:
\be\label{solution-su2}
\boxed{
K_{j_1,j_2,j_3}^{(\alpha)}= \frac{(-1)^{2\la+j_1+j_2}W(\la,j_1,\la, j_2;\la,j_3)}{\sqrt{N_{j_1} N_{j_2} N_{j_3}/N_0}},\qquad N_j=(-1)^{2(\la+j)}W(\la,j,\la, j;\la,0).
}
\ee
\be
K_{j_1,j_2,j_3}^{(\alpha)}=(-1)^{\frac{j_3-j_1-j_2}{2}}(2\alpha+1)^\frac12 [(2j_1+1)(2j_2+1)(2j_3+1)]^{\frac14} W(\la,j_1,\la, j_2;\la,j_3).\nonumber
\ee
We claim that the proposed structure constant $K_{j_1,j_2,j_3}^{(\alpha)}$ is symmetric under the permutations of its labels  and  satisfies the $SU(2)$-associativity equation \eqref{su2-crossing} or equivalently \eqref{su2-6j-crossing}. The symmetries of $K$ follow straightforwardly from the tetrahedral symmetry group of the Racah coefficient (more accurately, the $6j$-symbol) \eqref{6j-symmetry}. The proof of associativity follows just as straightforwardly from the Pentagon identity or the Biedenharn-Elliot identity of the Racah coefficient:
\bea\label{pentagon}
\sum_{j'}(2j'+1)
&& W(j_1,j_2,j_3,j_4;j,j') W(k_3,j_2,k_4,j_4;k_2,j')W(k_1,j_1,k_4,j';k_3,j_3)\nonumber\\
= && W(k_1,j_1,k_2,j_2;k_3,j)W(k_1,j,k_4,j_4;k_2,j_3).
\eea
The proof of this identity is discussed in  Appendix \ref{6j-symbol}.
Replacing all the $k$ type labels by $\la$ and using the normalizations in \eqref{solution-su2}, the above equation immediately reduces to the $SU(2)$-associativity equation.
In this way, we can construct infinitely many solutions to the associativity equation, each labeled by a representation $\la$.

In fact a more general class of solutions can be constructed in this way. Essentially by not requiring $\la$ to be a representation of $G$ but only requiring the operators acting on $\la$ be a representation of $G$.
The associativity of the operator product still guarantees that the ring coefficients are a solution to the required $G$-associativity equation. A closed form expression for the solutions of this general type  would be difficult to obtain.

The pentagon identity holds for any Lie group including the conformal group because it follows from a simple argument involving tensor product of four representations, as summarized in figure \ref{pentagon-fig}. Hence the $6j$-symbol for the conformal group, specialized to the above form gives a solution to the conformal crossing equation valued in the principal series. 
We have already outlined the procedure to go from the principal series representations to the physical representations by way of residue integral. 
So in effect, we have produced infinitely many solutions to the usual conformal crossing equations. In doing so, at no point, have we imposed the unitarity from the Lorentzian point of view. It would be interesting to see what constraints the Lorentzian unitarity imposes on these solutions.
As the solution to the crossing symmetry is essentially the $6j$-symbol for the conformal group, it would be important to compute it explicitly \cite{6j-symbol-work}. The asymptotic property of the structure constant controls the convergence of the integral in equation \eqref{assoc} which will in turn decide whether non-normalizable states, such as the identity, appear in the physical conformal spectrum. This question is also under investigation.

The easiest method for computing the conformal $6j$-symbol relies on version of the relation \eqref{CG-Racah} for the conformal group. Denoting the three point function by a trivalent vertex, the relation is succinctly expressed as
\begin{center}
\includegraphics[scale=0.4]{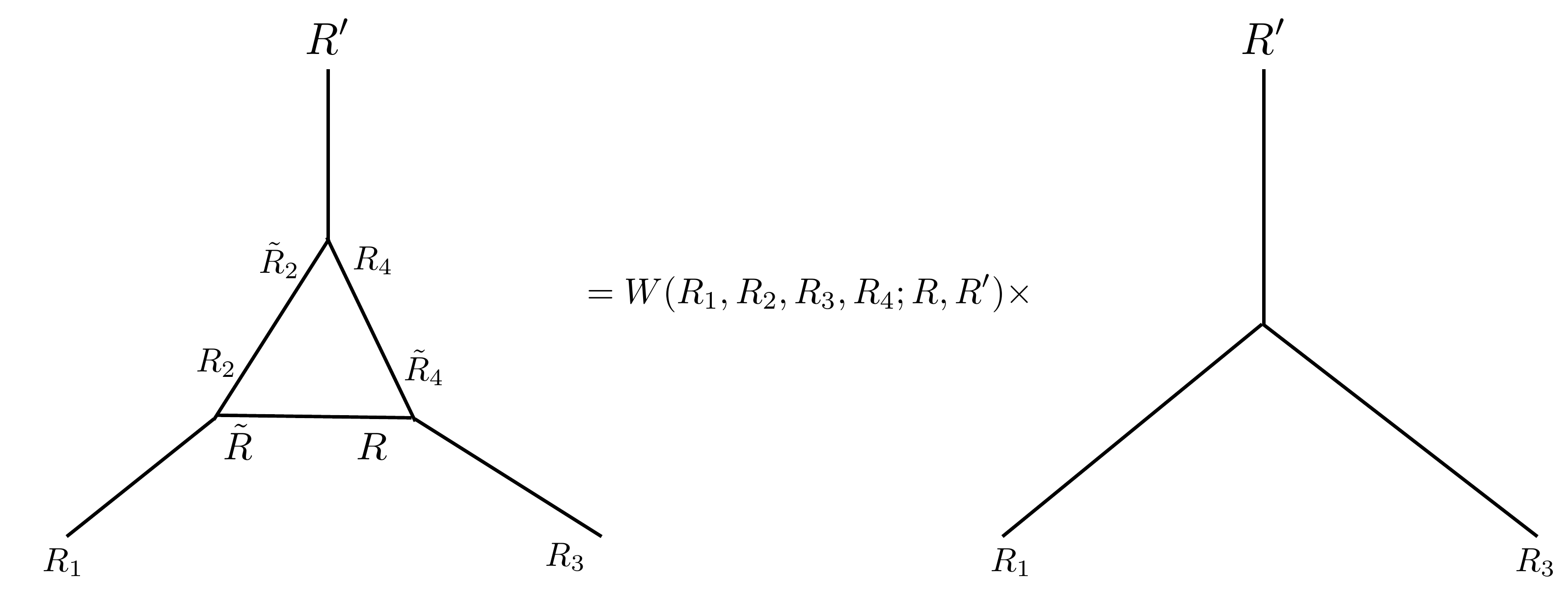}\label{conformal-6j}
\end{center}
From this equation we also see that the $6j$-symbol for the conformal group is invariant under exchanging on the of representations with its shadow representation. This means that in addition to the permutation symmetry of its labels, the proposed structure constant also has the symmetry
\be
K_{R_1,R_2,R_3}= K_{{\tilde R}_1,R_2,R_3}
\ee
as promised around equation \eqref{shadow-symmetry}.

Let us comment about the physical relevance of the solutions we have obtained. The solution for the $G$-crossing equation in terms of the $6j$ symbol appear as the boundary conformal theory of the $SU(2)_k$ WZW model \cite{Alekseev:1999bs, Runkel:1998he}. Here the group $G$ is taken to be the quantum group $SU(2)_q$, $q=\exp 2\pi i/(k+2)$. This is expected to be true for other groups as well.
The solution label $\la$ is the one that labels the conformal boundary conditions. The appearance of the quantum group in the context should not be surprising  as there is a well known connection between affine symmetries and quantum groups, first observed in \cite{AlvarezGaume:1988vr}. In fact the representation theory of the Virasoro algebra is also intimately linked to the representation theory of the quantum group $SL(2)_q$. 
Indeed the structure constants given in terms of the $6j$-symbol of $SL(2)_q$ appear as the structure constants for the FZZT brane, the famous conformal boundary condition of the Liouville theory \cite{Ponsot:2001ng}. In both these examples, the relevant structure constants are of the boundary theory and not of the bulk. This is because, on the boundary only one copy of the affine symmetry acts while on the bulk two copies, holomorphic and anti-holomorphic, of the same act. As a result the bulk structure constants obey the $G\times G$ crossing equation. The naive solution to this namely, the tensor product of the solutions to the $G$ crossing equation, doesn't work because the $G$-representations on holomorphic and anti-holomorphic side are typically entangled in a certain way as required by the modular invariance.

\section{Outlook}
In this paper we have looked at the conformal crossing equation from the viewpoint of conformal representation theory. As the conformal group is non-compact, its harmonic analysis plays an important role in this approach. Nevertheless, thanks to its well-developed theory, we are able to think of harmonic analysis as an additional feature for non-compact groups of what really is a problem for any Lie group. Using the crossing equation for $SU(2)$ group as a toy model, we have constructed infinitely many solutions to the crossing equation for any Lie group $G$ in terms of its $6j$-symbol. In particular, we have obtained infinitely many solutions to the conformal crossing equation. It would be interesting to investigate if they lead to physical unitary conformal field theories. 

What is the significance of the $G$-associativity equation for other groups $G$? Inspired by the kinematics of the AdS-CFT correspondence, we would like to postulate a correspondence between a ``$G$-Theory" \emph{i.e.} solution of $G$-associativity equation, and string theory on a certain  $G$-symmetric space. In the paper we have outlined how a topological quantum mechanical system with $G$ symmetry leads to the solution of $G$-associativity equation. 
The argument is equally valid if one replaces topological quantum mechanics with string theory. The relation between the space-time operator product expansion and world-sheet operator product expansion is explained in \cite{Aharony:2007rq}.
The example of the string theory of WZW model and its solution in terms of $6j$-symbol \cite{Alekseev:2000fd, Alekseev:2000wg}, mentioned at the end of the last section,  is indicative of such a correspondence. 
A connection between the de-Sitter gravity and a conformal theory of tempered representations has been pointed out in \cite{Chatterjee:2016ifv, Balasubramanian:2002zh}. This raises a tantalizing possibility, could the CMB 3-point functions be  $6j$-symbols?\footnote{We thank David Simmons-Duffin for asking this question.}
If such a generalized correspondence between a $G$-theory and a string theory exists then
we believe it would have most teeth in the case of non-compact groups because, in that case a $G$-theory can be expressed as a local quantum field theory on the coset  $G/MAN$ where $MAN$ is the distinguished parabolic subgroup of $G$ appearing in the harmonic analysis.
It would be a natural question is to develop a notion of large-$N$ expansion and match it with the string perturbation theory, \emph{\`a la} \cite{Heemskerk:2009pn, Aharony:2016dwx}.
Recent work on p-adic AdS/CFT correspondence \cite{Gubser:2016guj} could also fit in this framework by considering groups over p-adic numbers instead of over real/complex numbers. 

What can one say about solutions to conformal boundary conditions in higher dimensions? The local operators supported on the boundaries transform under a reduced symmetry group $SO(d,1)$. This can be thought of as the conformal group acting on the boundary. Hence the spectrum of boundary local operators and their three point function coefficients are subject to the boundary crossing equation \emph{i.e.} the crossing equation for the $SO(d,1)$ group. The solutions are found in the same way in terms of the $6j$-symbol of $SO(d,1)$. But the data, along with bulk-boundary structure constants, are also subject to an independent bulk-boundary crossing equation. We have found that our solutions are consistent with this additional constraint equation as well. This is the subject of an upcoming paper \cite{BCFT-solution}. We suspect that all the conformal defects fit in this framework consistently.

Our solutions could be particularly useful in the setting where the symmetries are believed to be most constraining. For example, the six dimensional $(0,2)$ superconformal theory is believed to be completely fixed, modulo a discrete label, by symmetries. If a physical solution is obtained for the  $(0,2)$ superconformal group crossing equation using our method  then in all likelihood it is \emph{the} $(0,2)$ superconformal theory. One can make a similar argument for the four dimensional ${\cal N}=4$ superconformal theory, in this case the conformal data is believed to depend on one additional continuous complex parameter. As our solutions are constructed from group theory, it is our hope that the unwieldy problem of conformal field theory classification gets reduced to relatively wieldy problem in  representation theory, at~least partially.

Recently, a connection has been found between $d$-dimensional Euclidean conformal theory data and scattering amplitudes in $d+2$-dimensional Lorentzian quantum field theory  \cite{Pasterski:2016qvg}. The conformal representations arising in this relation are not the physical ones but rather precisely the ones appearing in the harmonic analysis. These are also natural from our viewpoint. Hence, our solutions could be useful in describing  scattering in $d+2$ dimensions. As a $d+2$-dimensional theory is not just Lorentz invariant but rather Poincare invariant, one would need to impose the translational symmetry by hand. It would be nice to see how such a constraint can be naturally imposed on our solutions.

\section*{Acknowledgements}
We would like to thank Clay Cordova, Juan Maldacena, Pavel Putrov, Balt van Rees, David Simmons-Duffin and Douglas Stanford for illuminating discussions. We are also thankful to Shiraz Minwalla whose comments about conformal representation theory inspired this work. The author's  research is supported by the Roger Dashen Membership Fund and the National Science Foundation grant PHY-1314311.

\appendix
\section{The $SU(2)$ Racah coefficient}\label{6j-symbol}
The Racah coefficient is the recoupling coefficient for three angular momenta $j_1,j_2,j_3$ into $j_4$. Such a coupling can be achieved via two schemes: by first coupling $j_1$ and $j_2$ into $j$ and then coupling $j$ and $j_3$ into $j_4$ or by first coupling $j_2$ and $j_3$ into $j'$ and then coupling $j_1$ and $j'$ into $j_4$. They respectively yield
\bea\label{coupling-scheme}
|j_4,m_4; (j)\rangle &=& \sum_{m,m_3}  C^{j,j_3,j_4}_{m,m_3,m_4} \Big(\sum_{m_1,m_2} C^{j_1,j_2,j}_{m_1,m_2,m} |j_1,m_1\rangle |j_2,m_2\rangle \Big)  |j_3,m_3\rangle\nonumber\\
|j_4,m_4; (j')\rangle &=& \sum_{m_1,m'}  C^{j_1,j',j_4}_{m_1,m',m_4}   |j_1,m_1\rangle \Big(\sum_{m_2,m_3} C^{j_2,j_3,j'}_{m_2,m_3,m'} |j_2,m_2\rangle |j_3,m_3\rangle \Big).
\eea
 Here the coefficients $C$ are the Clebsch-Gordan coefficients. The bracketed label of $|j_4,m_4\rangle$ denotes the scheme used to construct it. Both coupling schemes construct orthonormal bases in the vector space $j_1\otimes j_2\otimes j_3$. Hence there must be a unitary matrix relating them. 
\be
|j_4,m_4; (j)\rangle = \sum_{j'} \sqrt{(2j+1)(2j'+1)}
 W(j_1,j_2,j_3,j_4;j,j')|j_4,m_4;(j')\rangle
\ee 
The matrix coefficient $W$ is known as  the Racah coefficient. Substituting \eqref{coupling-scheme} into this equation and using orthonormality of the basis vectors, we get a fundamental relation between the Racah coefficients and the Clebsch-Gordan coefficients.
\be\label{CG-Racah}
C^{j_1,j_2,j}_{m_1,m_2,m}C^{j,j_3,j_4}_{m,m_3,m_4} = \sum_{j'} \sqrt{(2j+1)(2j'+1)}
W(j_1,j_2,j_3,j_4;j,j')
\, C^{j_2,j_3,j'}_{m_2,m_3,m'}C^{j_1,j',j_4}_{m_1,m',m_4}.
\ee 
Using orthogonality and completeness of the Clebsch-Gordan coefficients \eqref{su2-ortho} further relations can be derived.
\bea\label{CG-Racah-relation}
&&\sum_{m_1,m_4}C^{j_1,j_2,j}_{m_1,m_2,m}C^{j,j_3,j_4}_{m,m_3,m_4} C^{j',j_1, j_4}_{m',m_1, m_4} = \sqrt{(2j+1)(2j'+1)}
W(j_1,j_2,j_3,j_4;j,j') C^{j_2,j_3,j'}_{m_2,m_3,m'}.\qquad \,\,\\
&&\sum_{m_1,m_4}C^{j_1,j_2,j}_{m_1,m_2,m}C^{j,j_3,j_4}_{m,m_3,m_4} C^{j',j_1, j_4}_{m',m_1, m_4} C^{j_2,j_3,j'}_{m_2,m_3,m'}= \sqrt{(2j+1)(2j'+1)}
W(j_1,j_2,j_3,j_4;j,j').\nonumber
\eea

\begin{figure*}[t]
\centering
\includegraphics[scale=0.3]{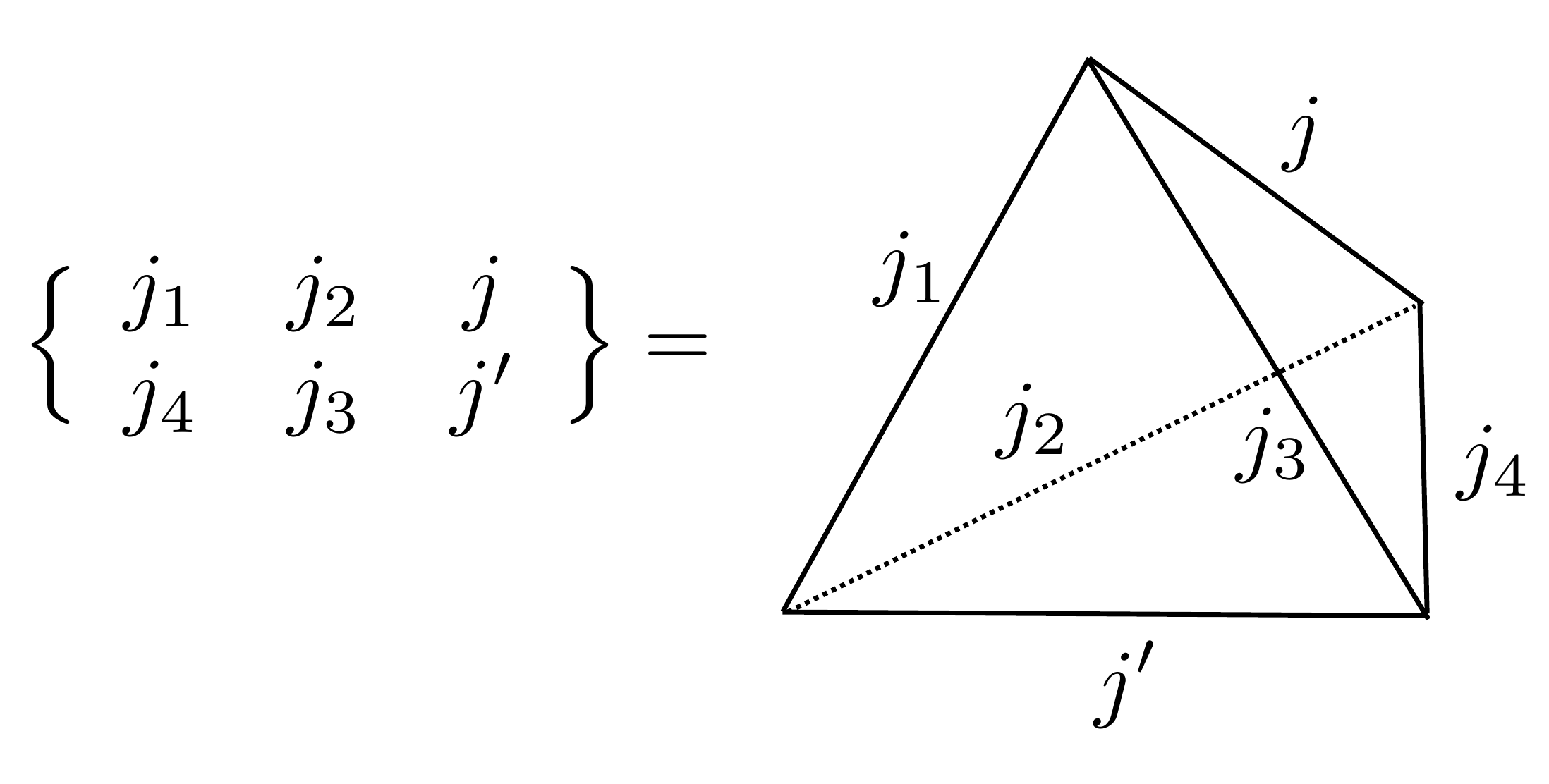}
\caption{Geometric interpretation of the $6j$-symbol as a tetrahedron.}
\label{tetrahedron}
\end{figure*}
It is sometimes convenient to use a more symmetric form of the Racah coefficient known as the $6j$-symbol. It is defined as,
\be\label{6j-symmetry}
\Big\{
\begin{array}{ccc}
j_1 & j_2 & j\\
j_4 & j_3 & j'
\end{array}
\Big\}
=(-1)^{j_1+j_2+j_3+j_4} W(j_1,j_2,j_3,j_4;j,j').
\ee
If we associate the six representations involved to six edges of a tetrahedron, as indicated in figure \ref{tetrahedron},  then the $6j$-symbol is invariant under the tetrahedral symmetry group.

\begin{figure*}[t]
\centering
\includegraphics[scale=0.3]{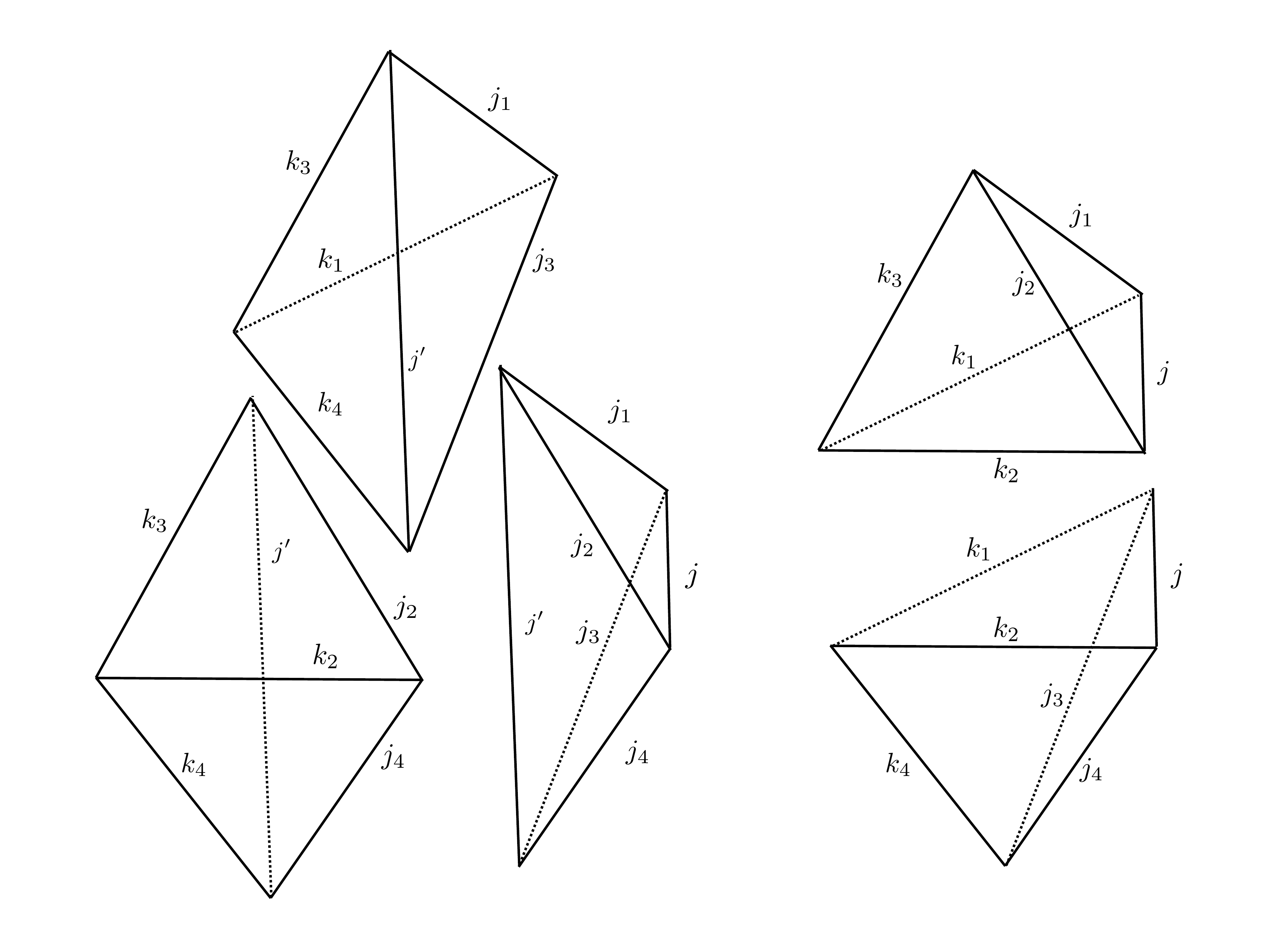}
\caption{Two triangulations of octahedron corresponding to the two sides of the pentagon identity.}
\label{pachner}
\end{figure*}

The Racah-coefficients obey a powerful identity known as the Pentagon identity or the Biedenharn-Elliot identity:
\bea\label{pentagon}
\sum_{j'}(2j'+1)
&& W(j_1,j_2,j_3,j_4;j,j') W(k_1,j_1,k_4,j';k_3,j_3) W(k_3,j_2,k_4,j_4;k_2,j')\nonumber\\
= && W(k_1,j,k_4,j_4;k_2,j_3) W(k_1,j_1,k_2,j_2;k_3,j).
\eea
From a geometric point of view, where we associate a tetrahedron to the Racah coefficient, the Pentagon identity is the equivalence between two ways of obtaining an octahedron: either  by gluing three tetrahedra along faces around a common edge  or  by gluing two tetrahedra along a face. This is shown in figure \ref{pachner}.
\begin{figure*}[t]
\centering
\includegraphics[scale=0.3]{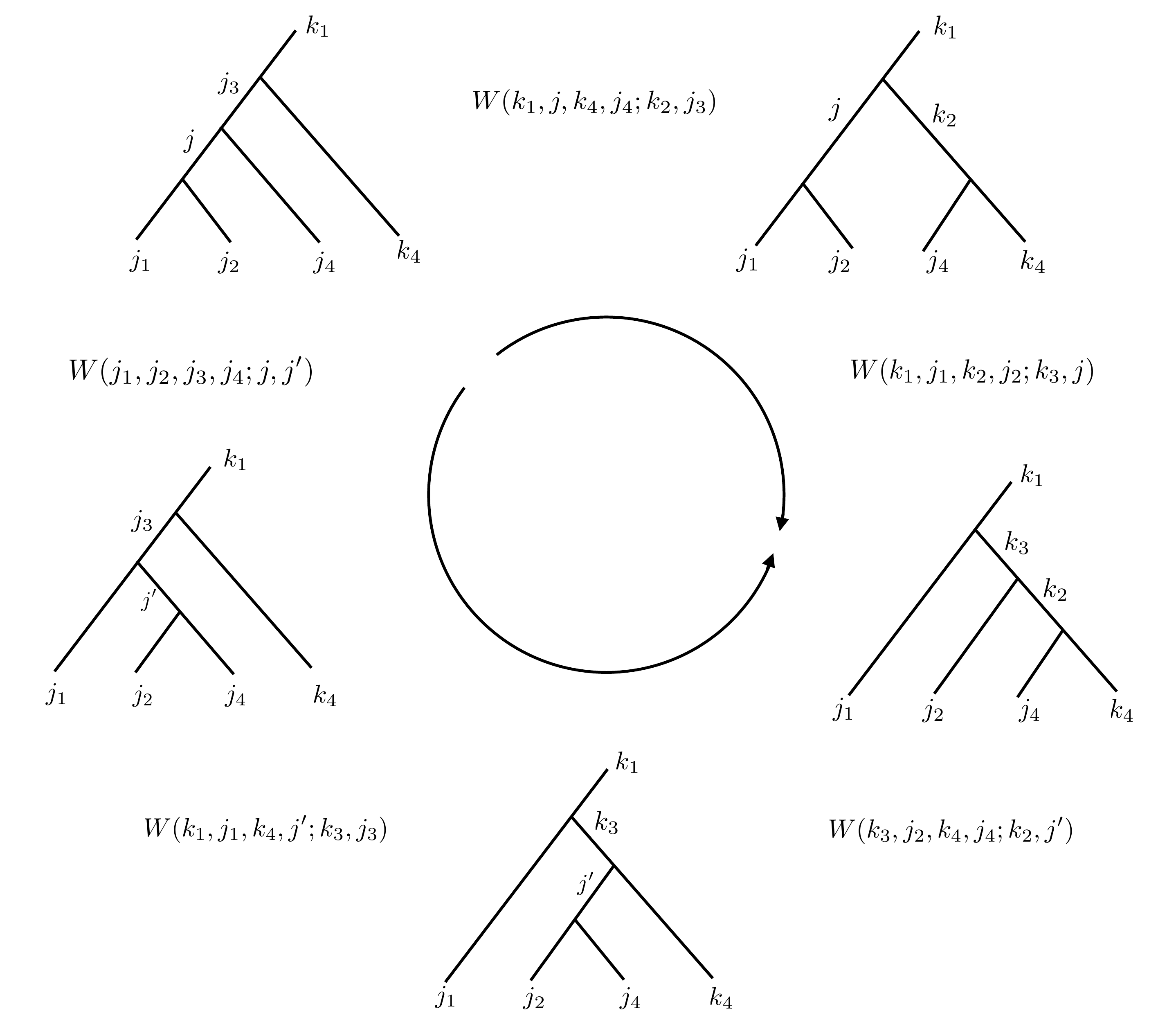}
\caption{The pentagon identity follows from considering the associativity of four angular momenta addition. The circular arcs denote the two sides of the pentagon identity.}
\label{pentagon-fig}
\end{figure*}
The pentagon identity follows when we consider coupling four angular momenta. Different ways of coupling them are sequentially related by multiplying Racah coefficients. Then the equivalence of two sets of ordered moves yields the pentagon identity. Again it is best to explain it in the graphical language, figure \ref{pentagon-fig}.
Due to the simple and robust origin of the pentagon identity, it holds for any group including the conformal group $SO(d+1,1)$.

\newpage

\bibliographystyle{JHEP_TD}
\bibliography{crossing}

\end{document}